\documentclass[rsi,amsmath,amssymb,reprint]{revtex4-1}

\usepackage{graphicx}
\usepackage{dcolumn}
\usepackage{bm}% bold math
\usepackage{subcaption}

\usepackage[utf8]{inputenc}
\usepackage[T1]{fontenc}
\usepackage{mathptmx}

\begin{document}

\preprint{RSI/Lapin}

\title[Study of Gasdynamic Electron Cyclotron Resonance Plasma Vacuum Ultraviolet Emission to Optimize Negative Hydrogen Ion Production Efficiency]{Study of Gasdynamic Electron Cyclotron Resonance Plasma \\ Vacuum Ultraviolet Emission to Optimize \\ Negative Hydrogen Ion Production Efficiency}

\author{R.L. Lapin}
 \email[]{lapin@ipfran.ru}
 \affiliation{ 
 Institute of Applied Physics, Russian Academy of Sciences, 603950 Nizhny Novgorod, Russia}
\author{V.A. Skalyga}
 \altaffiliation[Also at ]{Lobachevsky State University \\ of Nizhny Novgorod, 603155 Nizhny Novgorod, Russia.}
\author{I.V. Izotov}
\author{S.V. Razin}
\author{R.A. Shaposhnikov}
\author{S.S. Vybin}
\author{A.F. Bokhanov}
\author{M.Yu.~Kazakov}
 \affiliation{
 Institute of Applied Physics, Russian Academy of Sciences, 603950 Nizhny Novgorod, Russia}
\author{O.~Tarvainen}
 \affiliation{STFC ISIS Pulsed Spallation Neutron and Muon Facility, Rutherford Appleton Laboratory, Harwell, OX11 0QX, UK}
 \affiliation{University of Jyv{\"a}skyl{\"a}, Department of Physics, FI-40014 Jyv{\"a}skyl{\"a}, Finland}

\date{\today}

\begin{abstract}
Negative hydrogen ion sources are used as injectors into accelerators and drive the neutral beam heating in ITER. Certain processes in low-temperature hydrogen plasmas are accompanied by the emission of vacuum ultraviolet (VUV) emission. Studying the VUV radiation therefore provides volumetric rates of plasma-chemical processes and plasma parameters.

In the past we have used gasdynamic ECR discharge for volumetric negative ion production and investigated the dependencies between the extracted H$^-$ current density and various ion source parameters. It was shown that it is possible to reach up to 80 mA/cm$^2$ of negative ion current density with a two electrode extraction.

We report experimental studies on negative hydrogen ion production in a high-density gasdynamic ECR discharge plasma consisting of two simple mirror traps together with the results of VUV emission measurements. The VUV-power was measured in three ranges -- Ly$_\alpha$, Lyman band and molecular continuum -- varying the source control parameters near their optima for H$^-$ production. It was shown that the molecular continuum emission VUV power is the highest in the first chamber while Ly$_\alpha$ emission prevails in the second one. Modifications for the experimental scheme for further optimization of negative hydrogen ion production are suggested.

\end{abstract}

\maketitle %\maketitle must follow title, authors, abstract and \pacs

\section{Introduction}

Negative hydrogen ion sources are used as injectors into accelerators and drive the neutral beam heating in ITER \cite{bacal_2015}. Certain processes in low-temperature hydrogen plasmas, such as dissociation via repulsive excited states and molecular de-excitation populating high vibrational levels of the ground state molecules, are accompanied by the emission of electromagnetic radiation in vacuum ultraviolet (VUV) range \cite{komppula_2013}. Studying the VUV radiation therefore provides volumetric rates of plasma-chemical processes and plasma parameters. Furthermore, observing relative changes in the VUV emission intensity makes it possible to optimize the discharge conditions for efficient production of negative hydrogen ions.

Our previous works \cite{lapin_2017,lapin_2018} were dedicated to the research of using the gasdynamic ECR discharge for volumetric negative ion production and the investigation of dependencies between the extracted H$^-$ current density and various ion source parameters. We then made preliminary optimization of the extraction system consisting of two electrodes - plasma electrode and puller. The electron to ion current ratio was about 50 -- 85 in those experiments. It was shown that it is possible to reach up to 80 mA/cm$^2$ of negative ion current density.

Here we report experimental studies on negative hydrogen ion production in a high-density gasdynamic ECR discharge plasma together with the results of VUV emission measurements. We concentrate our attention on describing the experimental setup and reporting the measurement results whereas the interpretation will be communicated in the future.

\section{Experimental scheme}

\begin{figure*}
\includegraphics[width=1\linewidth]{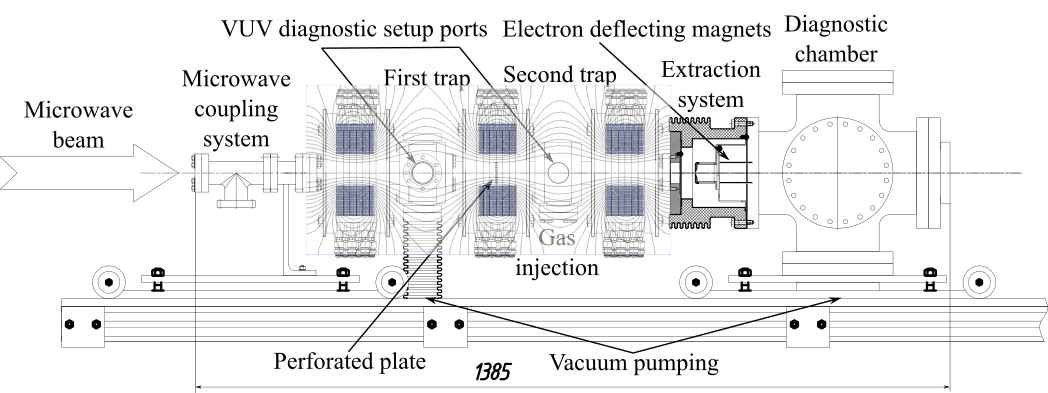}
\caption{\label{fig:scheme} Schematic view of the experimental setup depicting the magnetic field lines.}
\end{figure*}

The experimental facility is presented in Fig.\ref{fig:scheme}. The high density hydrogen plasma is sustained by a pulsed 37~GHz~/~100~kW gyrotron radiation in a two-stage magnetic system, consisting of two identical simple mirror traps. The microwave radiation was launched into the first chamber where the plasma is produced and sustained under the ECR condition. Dense plasma flux flows into the second trap through a perforated plate, which prevents the propagation of microwaves. The ion extraction was realized by two electrode system consisting of plasma and puller electrodes with the apertures of 1.8~mm and 3~mm, respectively, at a distance of 10~mm from each other. The plasma electrode was placed 103~mm downstream from the magnetic mirror. The neutral gas was injected into the second chamber in pulsed mode. The vacuum pumping was implemented from the first trap and the downstream diagnostic chamber.

The described configuration presumably helps to separate plasma volumes with ``hot'' electrons confined in the first trap and ``cold'' electrons confined in the second one. The plasma and vibrationally excited molecules flow into the second trap, where ``hot'' electrons lose their energy as a result of collisions with neutral gas molecules resulting in further ionization and excitation. The vibrationally excited molecules participate in the H$^-$ generation process as a result of dissociative ``cold'' electron attachment \cite{bacal_1979}.

\begin{figure}[h!]
\includegraphics[width=0.7\linewidth]{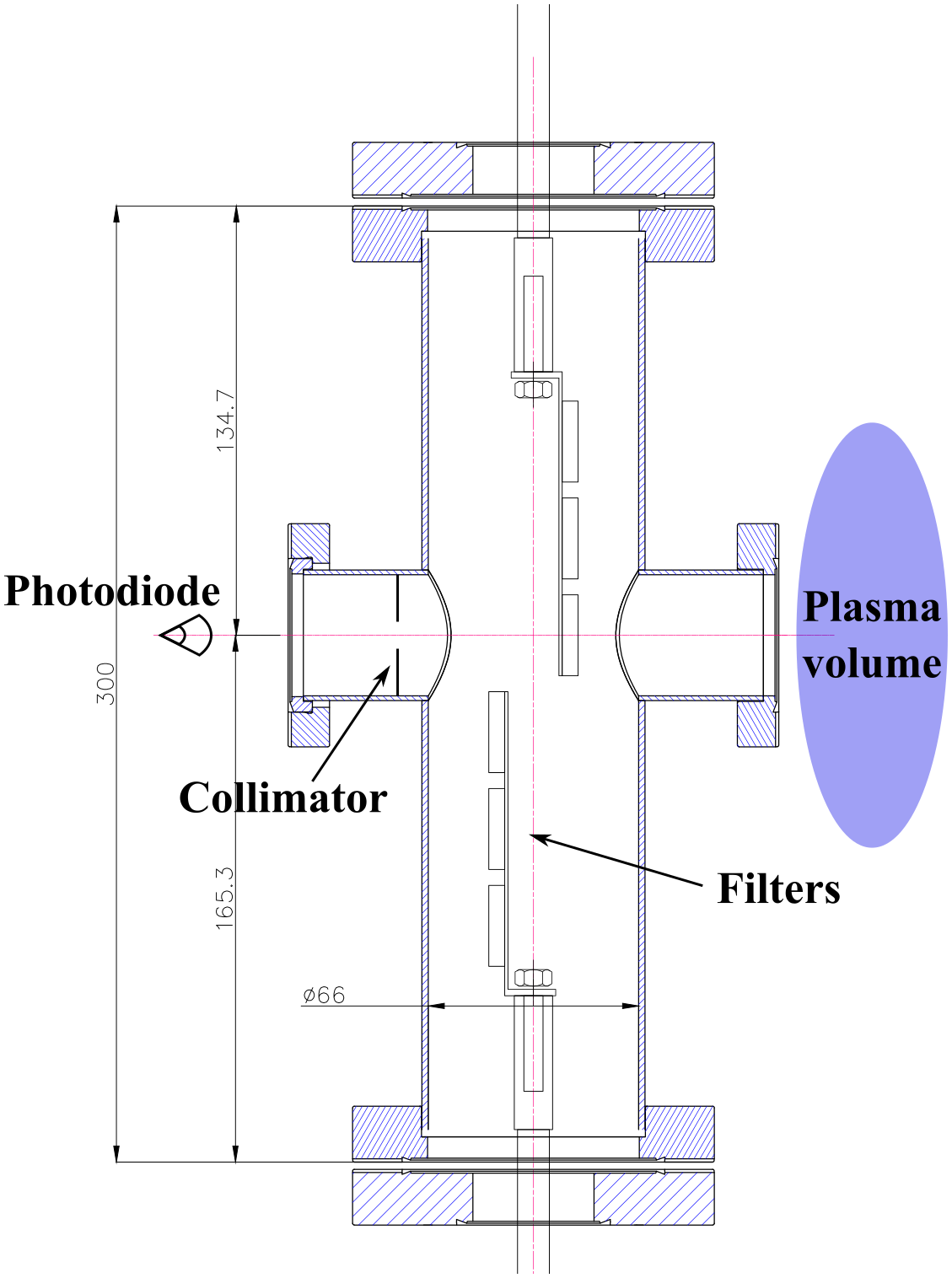}
\caption{\label{fig:setup} Scheme of the VUV-power density measurement.}
\end{figure}

The diagnostic system for the measurement of absolute power density of the VUV plasma emission is presented schematically in Fig.\ref{fig:setup}. Following \cite{komppula_2015}, it is based on three optical filters mounted on a movable steel tube in the vacuum chamber and a calibrated photodiode (IRD SXUV5). The filters correspond to atomic (122$\pm$10 nm –- Lyman-alpha line) and molecular emission (160$\pm$10 nm -– Lyman band and 180$\pm$20 nm –- molecular continuum) of hydrogen. The VUV diagnostic setup can be connected to both chambers (traps) viewing them through a radial port.

\begin{figure}[h!]
\includegraphics[width=1\linewidth]{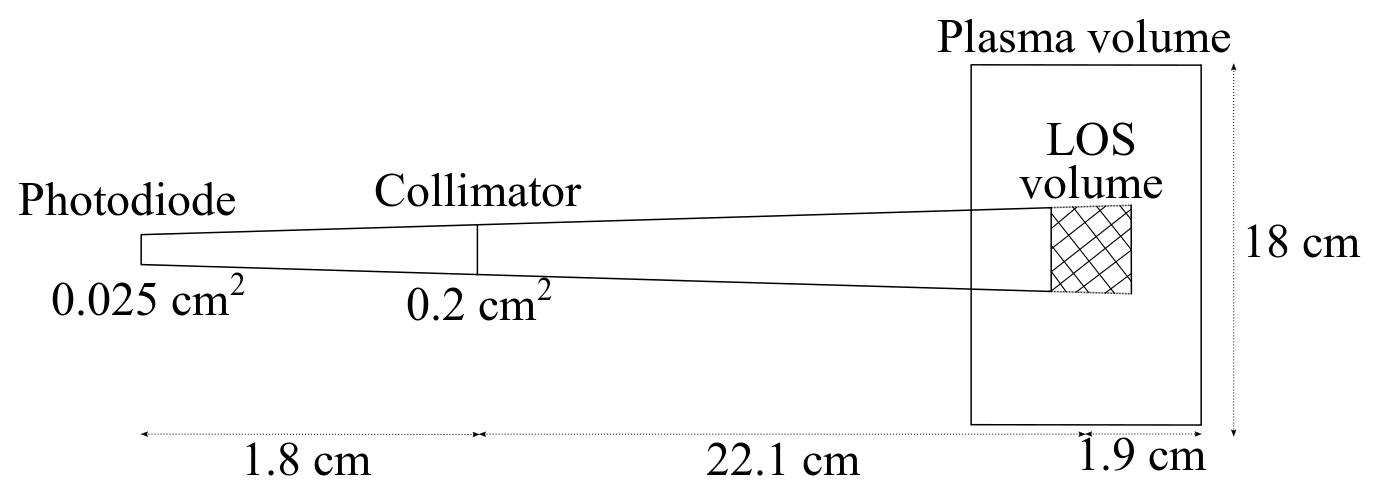}
\caption{\label{fig:plan} Dimensions required to calculate the VUV-power density from the measured photodiode current.}
\end{figure}

The absolute power density of the VUV plasma emission can be calculated from the measured current signal of the photodiode assuming isotropic emission and knowing the dimensions of the system, shown in Fig.\ref{fig:plan}.

%\begin{figure}[b]\setlength{\hfuzz}{1.1\columnwidth}
%	\centering
%	\begin{minipage}{0.4\textwidth}
%		\centering
%		\includegraphics[width=0.7\linewidth]{Fig2.eps}
%		\caption{\label{fig:setup} Scheme of the VUV-power density measurement.}
%	\end{minipage}
%	\begin{minipage}{0.59\textwidth}
%		\centering
%		\includegraphics[width=1\linewidth]{Fig3.eps}
%		\caption{\label{fig:plan} Dimensions required to calculate the VUV-power density from the measured photodiode current.}
%	\end{minipage}
%\end{figure}

The measured VUV-power from each chamber can be used for the evaluation of volumetric rates of various plasma processes, such as the production of high vibrational levels of the hydrogen molecules serving as precursors for the H$^-$ production through dissociative electron attachment.

\section{Results}

\begin{figure*}
	\includegraphics[width=0.7\linewidth]{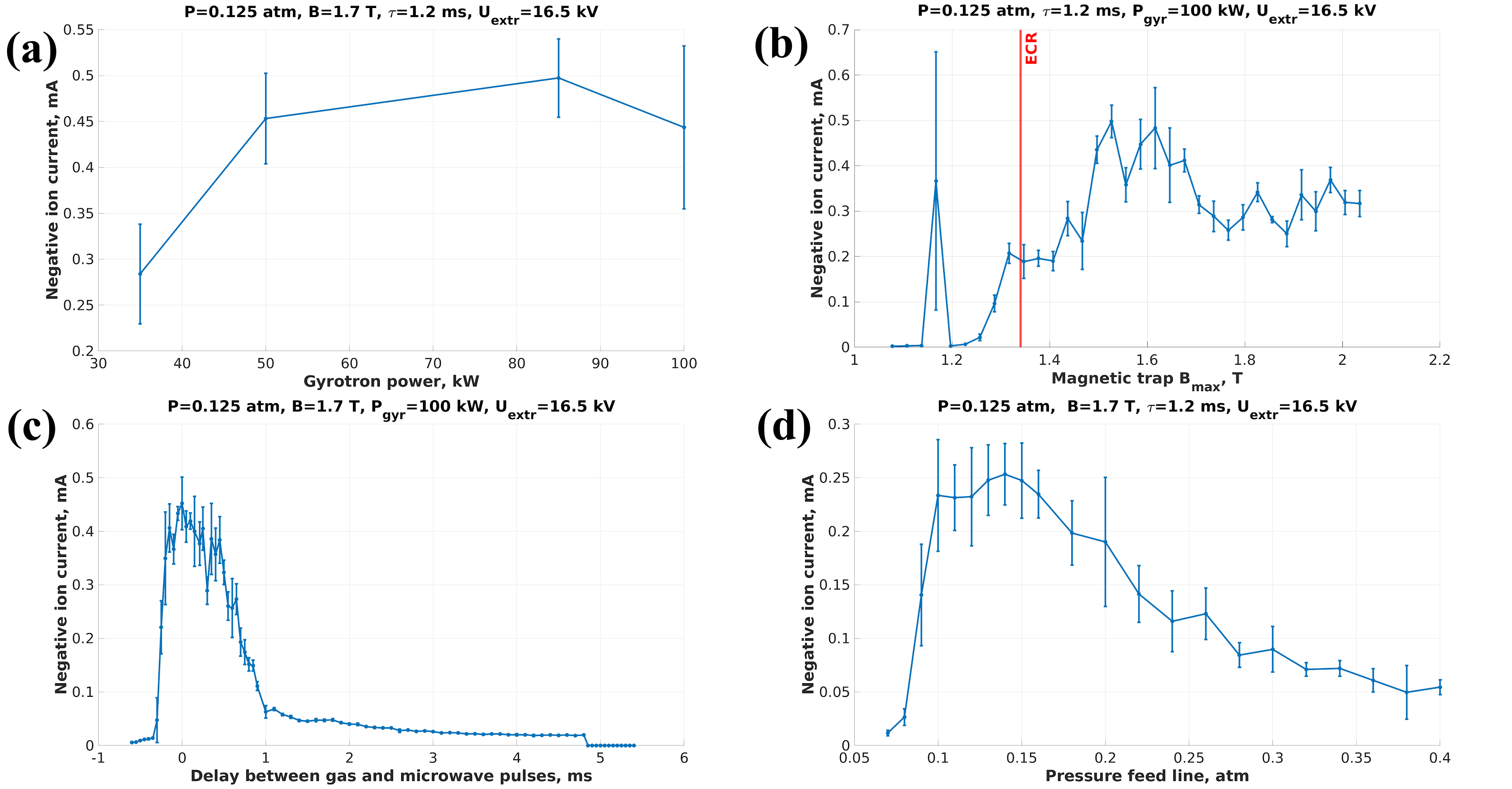}
	\caption{\label{fig:Hminus} The negative ion current dependence on (a) microwave power; (b) magnetic field; (c) the delay between the gas and microwave pulses; (d) gas feed line pressure.}
\end{figure*}

The maximum negative ion current was found using a Faraday cup and varying the temporal delay between the gas (4~ms) and microwave (1~ms) pulses, gas feed line pressure (both the delay and gas feed line pressure affect the pressure in the plasma chamber during the microwave pulse), the magnetic field in the two traps and the incident microwave power. The dependencies of the negative ion current on each of these parameters were investigated as presented in Figs \ref{fig:Hminus}a-d. The extraction voltage was set to 16.5~kV. It is believed that in this regime the negative ion current is limited by the negative ion density in the discharge, not by the space charge in the extraction system.

\begin{figure*}
	\includegraphics[width=0.7\linewidth]{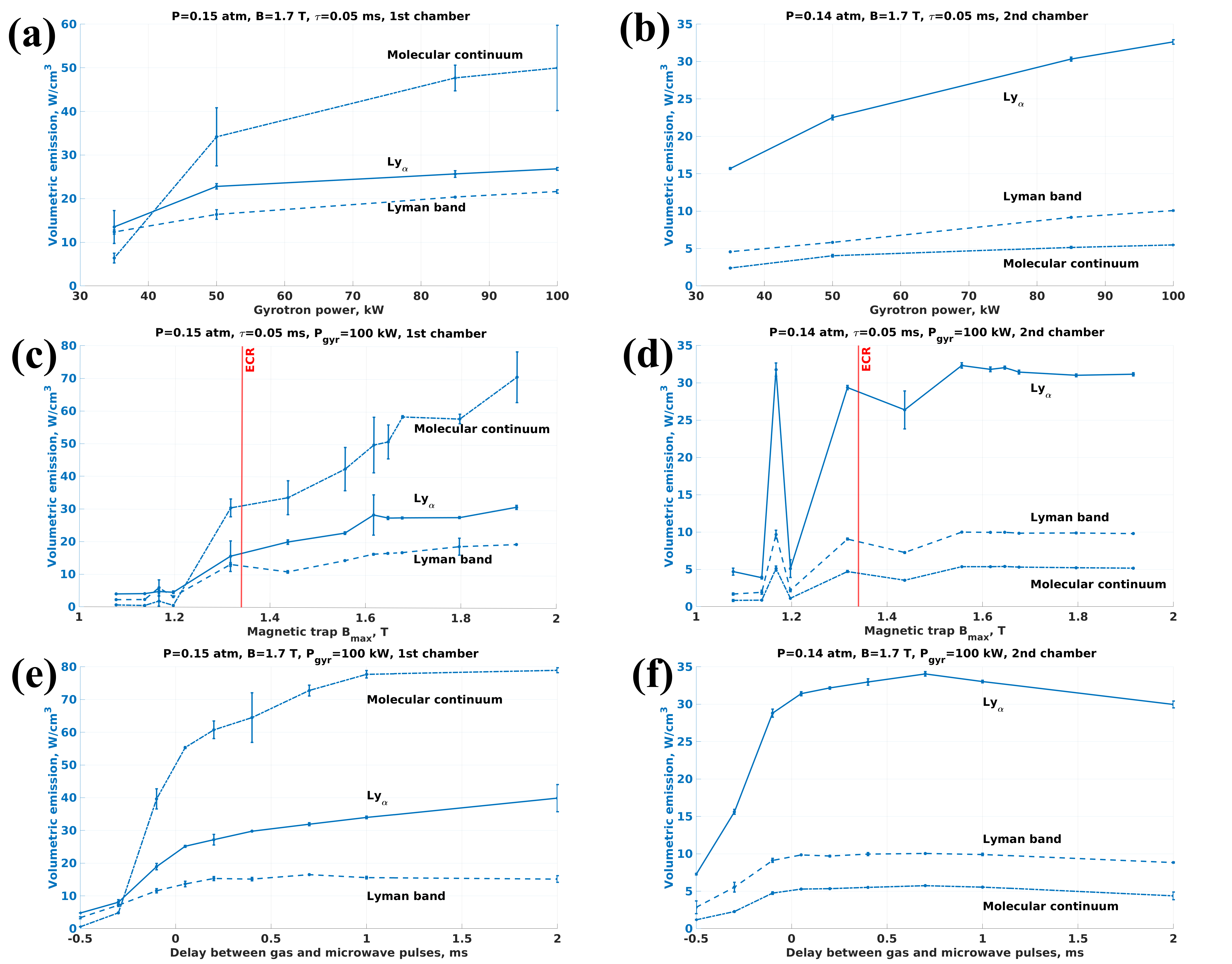}
	\caption{\label{fig:VUV} The dependence of VUV plasma emission on microwave power (a-b); on magnetic field (c-d); on the temporal delay between the gas pulse and the leading edge of the microwave pulse between (e-d) in both chambers respectively.}
\end{figure*}

The dependence of the H$^-$ current on gyrotron power (Fig.\ref{fig:Hminus}a) justifies the use of the gyrotron i.e. the negative ion current increases by a factor of two with the incident power in the range of 35 -- 85~kW. Fig.\ref{fig:Hminus}b shows that in the ECR regime (the resonance field B=1.34~T) it is possible to extract more negative ions than under off-resonance condition. The reason for the anomalous peak at B$\simeq$1.15~T is unclear. Figs \ref{fig:Hminus}c-d illustrate the same dependence of negative ion current on the neutral gas pressure during the discharge pulse, which is affected by the temporal delay between the gas and microwave pulses as well as gas feed line pressure. Negative delay means that the gas pulse was started after the microwave pulse.

The VUV emission from each trap was measured with the VUV diagnostic setup. Fig.\ref{fig:VUV} presents the dependencies of the VUV-power at different wavelengths on the experimental parameters.

The total VUV-power from the first chamber is higher by a factor of 3 in comparison to the second chamber. The main experimental observation is that the molecular continuum emission is dominant in the first chamber whereas the Ly$_\alpha$ emission has the highest power in the second one. Moreover, the values of Ly$_\alpha$-power are quite similar in both chambers. This implies that the atomic fraction is higher in the second trap, whereas in the first one the rates of molecular excitation processes are the dominant power dissipation mechanism. High atomic fraction in the second chamber is convergent with the molecular continuum emission leading to dissociation via the lowest (repulsive) triplet state being dominant in the primary trap and with the extremely high proton fraction of positive ion beams extracted from the SMIS-37 \cite{skalyga_2014}. Figs \ref{fig:VUV}a-b demonstrate that there is an increase of VUV emission with the increase of incident power. The VUV-power increases with the magnetic field in all three wavelength ranges in both chambers (Figs \ref{fig:VUV}c-d). In addition, there is a similar anomalous peak at B=1.15~T as was observed for the negative ion current. The VUV-power as a function of temporal delay (Figs \ref{fig:VUV}e-f) behaves differently in each chamber: the VUV emission increases monotonically in the first trap while it saturates in the second one. Both the VUV-emission power and negative ion currents increase with microwave power and B-field whereas their correlation as a function of the pressure is not apparent.

The measured data will be used for further modelling the negative hydrogen ion production.

\section{Conclusion}

The investigation of absolute VUV plasma emission power was performed for a 37.5~GHz ECR negative ion source. The dependencies of both negative ion current and VUV-power in three ranges –- Ly$_\alpha$, Lyman band and molecular continuum -- were measured as a function of operating parameters near the optimum for H$^-$ production. It was shown that the molecular continuum emission VUV power is the highest in the first chamber and Ly$_\alpha$ emission prevails in the second chamber. The changes of experimental scheme for further optimization of negative hydrogen ion production were suggested. The volumetric VUV-emission power in each wavelength range can be compared to those obtained with a 2.45~GHz microwave (ECR) source \cite{komppula_2015}. For SMIS-37 the volumetric power in the first trap was 15, 40 and 80~W/cm$^3$ (100~kW of total injected power) for Ly$_\alpha$, Lyman band and molecular continuum, respectively whereas in the microwave source these values are 0.05, 0.01 and 0.005~W/cm$^3$ with 600~W of total injected power. When normalized by the microwave power, i.e. to W/cm$^3$/kW, these numbers are rather similar, despite the two sources having different plasma volumes for example.

\begin{acknowledgments}
Work was supported by the grant of Russian Science Foundation, project \# 19-12-00377.
\end{acknowledgments}

\bibliography{H_minus}

\end{document}